\documentstyle[12pt,epsf,epsfig,cite]{article}

\newcommand{\beq}{\begin{equation}}
\newcommand{\eeq}{\end{equation}}
\newcommand{\bea}{\begin{eqnarray}}
\newcommand{\eea}{\end{eqnarray}}

\newcommand{\Vcb}{\left | {\rm V_{cb}}\right |}
\newcommand{\gevd}{{\rm GeV^2}}
\newcommand{\gevt}{{\rm GeV^3}}

\newcommand{\gevcd}{{\rm GeV/c^2}}

\newcommand{\gsim}{\lower.7ex\hbox{$
\;\stackrel{\textstyle>}{\sim}\;$}}
\newcommand{\lsim}{\lower.7ex\hbox{$
\;\stackrel{\textstyle<}{\sim}\;$}}

\def\lsim{\mathrel{\rlap{\lower3pt\hbox{\hskip0pt$\sim$}}
    \raise1pt\hbox{$<$}}}         
\def\gsim{\mathrel{\rlap{\lower4pt\hbox{\hskip1pt$\sim$}}
    \raise1pt\hbox{$>$}}}         

\newcommand{\bibit}[1]{\bibitem{#1}}

\newcommand{\GeV}{\,\mbox{GeV}}

\begin{document}

\begin{titlepage}
\pagestyle{empty}
\baselineskip=21pt
\rightline{CERN--TH/2002-290}
\rightline{ }
\vskip 0.05in
\begin{center}
{\large{\bf Heavy Quark Parameters and \boldmath $\,|V_{cb}|\,$ from\\ 
Spectral Moments in Semileptonic  $B$ Decays
}}
\end{center}
\begin{center}
\vskip 0.08in
{{\bf Marco~Battaglia}$^1$,
{\bf Marta~Calvi}$^2$,
{\bf Paolo~Gambino}$^1$,
{\bf Arantza~Oyanguren}$^3$,
{\bf Patrick~Roudeau}$^4$,
{\bf Laura~Salmi}$^5$,
{\bf Jose~Salt}$^3$,
{\bf Achille~Stocchi}$^4$,
{\bf Nikolai~Uraltsev}$^{2,6,7}$
}
\vskip 0.08in
{\it
$^1${CERN, Geneva, Switzerland}\\
$^2${Universita' degli Studi di Milano-Bicocca
and INFN. Sezione di Milano (Italy)}\\
$^3${Instituto de Fisica Corpuscular, Universitat Valencia (Spain)}\\
$^4${LAL Orsay (France)}\\
$^5${Helsinki Institute of Physics (Finland)}\\
$^6${Dept. of Physics, University of Notre 
Dame du Lac, Notre Dame, IN 46556, USA}\\ 
$^7${St.\,Petersburg Nuclear Physics Institute, Gatchina, St.\,Petersburg 
188300, Russia}
}
\vskip 0.35in
{\bf Abstract}
\end{center}
\baselineskip=18pt \noindent

We extract the heavy quark masses and 
non-perturbative parameters from the {\sc Delphi} preliminary measurements 
of the first three moments of the charged lepton energy and hadronic mass 
distributions in semileptonic $B$ decays, using a multi-parameter fit. 
We adopt two formalisms, one of which 
does not rely on a $1/m_c$ expansion and makes use of running quark masses. The data 
are consistent and the level of accuracy of the experimental inputs largely 
determines the present sensitivity. The results allow to improve on the uncertainty 
in the extraction of $|V_{cb}|$.

\vfill
\vskip 0.15in
\leftline{CERN--TH/2002-290}
\leftline{23 October 2002}
\end{titlepage}
\baselineskip=18pt

\section{Introduction}

The Operator Product Expansion (OPE) represents a foundation for extracting the 
$|V_{ub}|$ and $|V_{cb}|$ elements of the CKM mixing matrix from inclusive 
semileptonic (s.l.)\ $B$ decays. In this framework, the decay width is expressed in 
terms of quark masses, and non-perturbative effects are described by expectation values 
of heavy quark operators, some of which are presently poorly known.
The experimental accuracy already achieved, and that expected from the large data 
sets recorded by the $B$-factories, makes the ensuing theory uncertainty a serious 
limitation. Extracting the heavy quark masses and the non-perturbative parameters, 
arising from the $1/m_b^2$ and $1/m_b^3$ corrections, directly from the data has 
therefore become a key issue. 
There have already been $|V_{cb}|$ determinations from the
first moment of distributions in s.l.\ and $b \to X_s \gamma$ decays, and the 
$1/m_b^3$ corrections, estimated from parameter ranges, have been found to represent 
an important source of uncertainty~\cite{Briere:2002hw}. 
These ranges, based on dimensional arguments, are affected by a considerable 
degree of arbitrariness. 

In order to circumvent these problems, we introduce in this Letter a multi-parameter 
fit to determine the relevant $1/m_b^2$ and $1/m_b^3$ parameters, together with the 
heavy quark masses, from the first three 
moments of the leptonic energy and hadronic mass spectra in s.l.\ $B$ decays. 
Results are based on preliminary data obtained by the {\sc Delphi} 
Collaboration. Moments are measured without cuts on the lepton energy in the 
$B$ rest frame. We consider two formalisms, one of which is new and 
relies on fewer theoretical assumptions.
The use of higher moments guarantees a sensitivity to these parameters and the 
simultaneous use of the hadronic and leptonic spectra ensures that a larger number of 
parameters can be kept free in the fit. 
We discuss the results both in terms of the extraction of the parameters and the 
implications for $|V_{cb}|$, and as a consistency check of the underlying theoretical 
assumptions.


\section{Extracting Non-Perturbative Parameters}

The moments of the hadronic and leptonic spectra in s.l.\ $B$ decays have recently been 
measured by several experiments \cite{Cronin-Hennessy:2001fk,Briere:2002hw,delphi_mx,
delphi_el,babar}. 
We consider here moments of the charged 
lepton energy distribution defined as
\beq
M_1(E_\ell) = \frac1{\Gamma}\,\int d E_\ell\, E_\ell
\frac{d\Gamma}{dE_\ell}; \quad\quad
M_n(E_\ell) = \frac1{\Gamma}\,\int d E_\ell\,
\left(E_\ell-M_1(E_\ell)
\right)^n  \frac{d\Gamma}{dE_\ell}\ \ \ (n>1),
\label{eq:01}
\eeq
and moments of the distribution of $M_X$, the invariant hadronic mass,
\beq
M_1(M_X)\!=\!\frac1{\Gamma}\,\int dM_X^2\, (M_X^2\!-\!\bar{M}_D^2)\frac{d\Gamma}{dM_X^2}
; \quad\quad
M_n(M_X)\!=\!\frac1{\Gamma}\,\int dM_X^2\, (M_X^2\!-\!\langle M_X^2 \rangle)^n 
\frac{d\Gamma}{dM_X^2}  \ \ (n>1),
\label{eq:02}
\eeq
where $\bar{M}_D=1.973$~GeV is the spin averaged $D$ meson mass and no cut on the 
charged lepton energy is assumed. 

The theoretical framework to interpret these data has long been known 
and it is based on the OPE. Different implementations exist, depending on the way the 
quark masses are treated. For instance, 
the $m_b$ and $m_c$ masses can be taken as independent 
parameters or subject to a constraint on $m_b\!-\!m_c$, imposed from the measured 
$B^{(*)}$ and $D^{(*)}$ meson masses. 
The second choice introduces a $1/m_c$ expansion.
Another option concerns the normalization scheme used for quark masses and 
nonperturbative parameters. One approach is to use short-distance masses, such as 
the low-scale running masses. Alternatively, the pole mass scheme can be used.

The OPE expresses lepton moments through quark masses as a double
expansion in $\alpha_s$ and $1/m_b$:
\bea
M_n(E_\ell) = \left(
\frac{m_b}{2}\right)^n  \left[ 
\varphi_n(r)  + \bar{a}_n(r) \,\frac{\alpha_s}{\pi} 
+\bar{b}_n(r)\,\frac{\mu_\pi^2}{m_b^2}
+\bar{c}_n(r)\,\frac{\mu_G^2}{m_b^2}
+\bar{d}_n(r)\,\frac{\rho_D^3}{m_b^3}
+\bar{s}_n(r)\,\frac{\rho_{LS}^3}{m_b^3} + ...
\right] ,
\label{eq:03}
\eea
where $r=(m_c/m_b)^2$. The higher coefficient functions $\bar{b}(r)$, 
$\bar{c}(r)$, ... are also 
perturbative series in $\alpha_s$. The expectation values of only two operators 
contribute to the $1/m_b^3$ corrections: the Darwin term $\rho_D^3$ 
and the spin-orbital term $\rho_{LS}^3$.
Due to the kinematic definition of the hadronic invariant mass $M_X^2$, the
general expression for the hadronic moments includes $M_B$ explicitly:
\bea
\nonumber
{M}_n(M_X)&=&m_b^{2n}\,\sum_{l=0} \;
\left[\frac{M_B-m_b}{m_b}\right]^l
\left\{E_{nl}(r) + a_{nl}(r)\frac{\alpha_s}{\pi}
+b_{nl}(r)\frac{\mu_\pi^2}{m_b^2} +
c_{nl}(r)\frac{\mu_G^2}{m_b^2} \right.\\
&& \left. \hspace{5.3cm}
+d_{nl}(r)\frac{\rho_D^3}{m_b^3} +
s_{nl}(r)\frac{\rho_{LS}^3}{m_b^3} + ...\right\}.
\label{eq:04}
\eea

It is possible to re-express the heavy quark masses, $m_Q$, in the above equations, 
in terms of the meson masses, $M_{H_Q}$, through the relation~\cite{Bigi:ga}: 
\beq
\label{eq:05}
M_{H_Q} = m_Q+ \bar{\Lambda} + 
\frac{\mu_\pi^2-\mu_G^2}{2 m_Q} + 
\frac{\rho_D^3 + \rho_{LS}^3 - \rho_{nl}^3}
{4 m_Q^2} + {\cal{O}}\left(\frac{1}{m_Q^3}\right).
\eeq
The use of these expressions
introduces an explicit dependence of the non-local correlators contributing to
$\rho_{nl}^3$. In the notation of~\cite{mb3}, $\rho_{nl}^3$ 
corresponds to linear combinations of ${\cal T}_{1-4}$.

Here, we employ the following two formalisms. The first one is based on 
the kinetic running masses, $m_Q(\mu)$, and non-perturbative parameters, 
introduced in~\cite{kinmass}. No charm mass expansion is assumed. 
The second formalism employs quark pole masses and the $B^{(*)}$ and $D^{(*)}$ 
meson mass relations. It represents a useful reference, as it has been already 
adopted in several studies. 

Contributions through $O(\alpha_s^2 \beta_0)$ \cite{alphas,alphas2b0}
and $O(1/m_b^3)$ \cite{bigi1,voloshin,FLS,Falk:1997jq} to the moments are available. 
Depending on the formulation adopted, the number of parameters involved at this 
order ranges from six to nine. Some of these parameters, like $m_b$ and 
$\lambda_2 \simeq \mu_G^2/3$, are relatively well known. Others, notably those which 
appear at $O(1/m_b^3)$, are virtually unknown.

\subsection{The $m_b(\mu)$, $m_c(\mu)$ and $\mu_{\pi}^2(\mu)$ Formalism}
The running kinetic quark masses $m_b(\mu)$ and $m_c(\mu)$ are considered here
as two independent parameters.
Apart from $\mu_\pi^2(\mu)$ and $\mu_G^2(\mu)$, defined as 
expectation values in the actual $B$ meson, there are two 
$1/m_b^3$ parameters, $\rho_D^3$ and $\rho_{LS}^3$.
The effect of $\rho_{LS}^3$ turns out to be numerically small.
In Eqs.~(\ref{eq:03}) and (\ref{eq:04}) the mass ratio $r$ is given by 
$(m_c(\mu)/m_b(\mu))^2$, and the $b$ quark mass is understood as $m_b(\mu)$.
The perturbative coefficients additionally depend on $\mu/m_b$ and the mass 
normalization scale $\mu$ is set at $\mu=1\GeV$.
The functions $\varphi_n$ in Eq.~(\ref{eq:03}) are well-known parton expressions. 
The relevant coefficients are given in Table~\ref{tab:1}, for the central values of 
$m_b(1~{\rm{GeV}})$ = 4.6~GeV and $r \simeq 0.06$ obtained in our fit. Although we
quote only the leading-order perturbative coefficients, we also include second-order 
BLM corrections in the analysis. Detailed expressions for the coefficients 
will be presented elsewhere.

\begin{table}
\caption{\sl Numerical values of the coefficients in Eq.(\ref{eq:03}) 
evaluated at $r$=0.06 and $m_b(1\,{\rm GeV})= 4.6\,{\rm GeV}$.}
\begin{center}
\begin{tabular}{|l|c|c|c|c|c|c|}
\hline 
 & $\varphi_n$ & $\bar{a}_n$ & $\bar{b}_n$ & $\bar{c}_n$ & $\bar{d}_n$ & 
$\bar{s}_n$ \\
\hline
$M_1(E_\ell)$ & 0.6173 & 0.015  &0.31 &-0.73 &-3.7 &0.2 \\
$M_2(E_\ell)\,(\times 10)$ & 0.3476 & 0.026  &1.7 &-1.0 & -10.2& -0.9\\
$M_3(E_\ell)\,(\times 10^2) $ & -0.3410 &0.066 & 3.4 & 1.3&-23& -4.2\\
\hline
\end{tabular}
\end{center}
\label{tab:1}
\end{table}

In the case of hadronic moments, we discard in Eq.~(\ref{eq:04})  coefficients 
$b_{nl}$, $c_{nl}$ with $l\!>\!1$, and $d_{nl}$, $s_{nl}$ with $l\!>\!0$. 
The only non-vanishing $E_{i0}$ coefficient is 
$E_{10} = r - \bar{M}_D^2/m_b^2$. The value of the other coefficients, at 
$r=0.06$, are listed in Table~\ref{tab:2}. Here we consider only
${\cal O}(\alpha_s)$ corrections and evaluate them using $\alpha_s=0.3$.

\begin{table}
\caption{\sl Numerical values of the coefficients in Eq.(\ref{eq:04}) evaluated 
at $r$=0.06 and $m_b(1~{\rm{GeV}}) = 4.6\,{\rm GeV}$.}
\begin{center}
\begin{tabular}{|l|c|c|c|c|c|c|c|c|c|c|c|}
\hline 
 $i$ & $E_{i1}$ & $E_{i2}$ & $E_{i3}$ & $a_{i0}$&$a_{i1}$ & 
$b_{i0}$ & $b_{i1}$
 &$c_{i0}$ & $c_{i1}$& $d_{i0}$ & $s_{i0}$  \\
\hline
 $1$ &0.839 &1 &0 &  0.029  &0.013 & -0.58 &-0.58 &0.31 &0.87&3.2 
&-0.4  \\
 $2$ &0 &0.021 &0 &- 0.001  &-0.002& 0.16 &0.34 &0 &-0.05 &-0.8 &0.05 
\\
 $3$ &0 & 0&-0.0011 & 0.0018 &  0.0013  &0 &0.034 &0 &0 &0.15  &0\\
\hline
\end{tabular}
\end{center}
\label{tab:2}
\end{table}

\subsection{The $\bar \Lambda$ and $\lambda_1$ Formalism}
This widely used scheme results from the combination of the OPE with the HQET. 
Following the notation of Ref.~\cite{Falk:1997jq}, the moments are 
 expressed in the following general form:
\bea\label{eq:06}
M_n&=& M_B^k \left[a_0  +
a_1 \frac{\alpha_s(\bar{M}_B)}{\pi} 
 +a_2 \beta_0 \frac{\alpha_s^2}{\pi^2} 
+ b_1 \frac{ \bar{\Lambda}} {{\bar{M}_B}} 
+ b_2 \frac{\alpha_s}{\pi}\frac{ \bar{\Lambda}} {{\bar{M}_B}} 
+ \frac{c_1\,\lambda_1 + 
     c_2\, \lambda_2 +      c_3\,{ \bar{\Lambda}}^2}
     {{ {\bar{M}_B}}^2}  
\right. \nonumber\\
&& \left.+ \frac1{\bar{M}_B^3} \left(d_1\,\lambda_1\bar{\Lambda}
      +      d_2\, \lambda_2\bar{\Lambda} +  d_3\,{ \bar{\Lambda}}^3  + 
     d_4\, \rho_1 +d_5\, \rho_2 + \sum_{i=1,4} d_{5+i}{{\cal T}_i}
\right)+ O\left( \frac{\Lambda^4_{QCD}}{{m}^4_Q}\right)\right],
\eea
where $k=n$ and $k=2n$ for leptonic and hadronic moments,
respectively, while $a_0=0$ for hadronic moments. $\bar{M}_B=5.3135$~GeV is the
spin-averaged $B$ meson mass.
The second order BLM corrections\footnote{The terms $O(\alpha_s^2 \beta_0)$ and
$O(\alpha_s \bar{\Lambda})$ are not available for 
the third hadronic moment. In our analysis we employ an estimate and the
related uncertainty is included in the fit.}
are expressed  in terms of $\beta_0=11-2/3 n_f$, where we take $n_f=3$.
The coefficients $a_i,b_i,c_i,d_i$ are given in Table~\ref{tab:3} for the first three
leptonic, $M_{1,2,3}(E_{\ell})$, and hadronic moments, $M_{1,2,3}(M_X)$.
In the leptonic case the $a_1$  coefficients agree with ref.~\cite{voloshin}, while the 
coefficients of the first two hadronic moments agree with \cite{FLS,Falk:1997jq}.
Details of the derivation will be presented elsewhere.

The non-perturbative parameters in Eq.(\ref{eq:06}) are related to those in Section~2.1 
by the following relations, valid up to ${\cal{O}}(\alpha_s)$:
\beq
\mu_{\pi}^2 =-\lambda_1 - \frac{{\cal T}_1 + 3 {\cal T}_2}{m_b};
\quad
\mu_G^2 =3\lambda_2 + \frac{{\cal T}_3 + 3 {\cal T}_4}{m_b};
\quad 
\rho_D^3=\rho_1\, ; 
\quad
 \rho_{LS}^3=3 \rho_2\,.
\eeq 
Perturbative corrections introduce a significant numerical difference 
between the parameters in the two schemes. At $\mu=1~\GeV$:
\beq
\bar{\Lambda} \simeq M_B - m_b{\rm{(1~GeV)}} - \frac{\mu_{\pi}^2 - \mu_G^2}{2 m_b} - 
0.26~{\rm{GeV}}~;
\quad \quad
-\lambda_1 \simeq \mu_{\pi}^2{\rm{(1~GeV)}} -
0.17~{\rm{GeV}}^2\,.
\eeq

\begin{table}
\caption{\sl Numerical values of the coefficients in Eq.(\ref{eq:06}).}
\begin{center}
\begin{tabular}{|l|c|c|c|c|c|c|c|c|c|}
\hline 
 & $a_0$ & $a_1$ & $a_2$ & $b_1$ & $b_2$ & $c_1$ & $c_2$ & $c_3$ & \\
 & $d_1$ & $d_2$ & $d_3$ & $d_4$ & $d_5$ & $d_6$ & $d_7$ & $d_8$ & $d_9$ 
\\
\hline
$M_1(E_\ell)$ $(\times 10)$ &2.708 &-0.004 &-0.10 & -0.548& -0.15&-3.99 &-9.77 &
 -0.77& \\
 &-10.1 & -15.3&-1.2 &-9.7 &3.1 &-3.9 &4.1 &-2.0 &9.8 \\ \hline
$M_2(E_\ell)$ $(\times 10^2)$ &0.710 &-0.096 & -0.18&-0.535 &-0.10 &-4.32 &-5.75 
&-0.35 & \\
 &-7.2 & -8.1& -0.2&-19.7 & -5.4&1.3 & 10.2& -0.2&5.7\\ \hline
$M_3(E_\ell)$ $(\times 10^3)$ &-0.257 &-0.014 &0.03 & -0.017&-0.01
 &-2.14 & 2.88&0.20 & \\
&0.5 &5.6 & 0.4&-28.3 &-11.4 &5.2 & 9.6& 1.0&-2.9\\ \hline
$M_1(M_X)$    &0 &0.052 &0.096  &0.225 &0.10 &1.04 &-0.31  & 0.28& \\
    &2.2  &2.4 &0.3 & 2.3&-1.2 &1.6 &0.8 &1.5  & 0.4 \\
 \hline
$M_2(M_X)$ $(\times 10)$ &0 &0.054 &0.078 &0 &0.14 &-1.40 & 0& 0.11& \\
& -1.6& -1.6&0.2 &-8.7 &2.4 &-1.4 & -4.2 & 0 & 0\\
 \hline
$M_3(M_X)$ $(\times 10^2)$     &0 &0.106 &-- &0 &-- &0 &0 &0 & \\
    &-2.05 &0 &-0.03 &14.45&0 &0 &0  &0 &0\\
\hline
\end{tabular}
\end{center}
\label{tab:3}
\end{table}

A well known problem of this formalism is the instability of the
perturbative series, due to the use of the pole quark masses. 
Large higher order corrections  are however expected to cancel in the relation 
between physical observables, as long as all observables involved 
in the analysis are computed at the same order in 
$\alpha_s$~\cite{Bigi:ga,Uraltsev:1995jx}. 
We also note that, 
as a consequence of the HQET mass relations for the mesons, the intrinsic 
expansion parameter 
in Eq.(\ref{eq:06}) is $1/M_D$, rather than $1/M_B$. The convergence
of this expansion has been questioned, in view of 
indications~\cite{Uraltsev:2001ih,lattice} that the matrix elements 
${\cal T}_i$  of some non-local operators could be larger than that 
expected from dimensional estimates. 

\section{Fits and Results}



This analysis is based on the preliminary {\sc Delphi} 
measurements~\cite{delphi_mx,delphi_el} of 
the first three moments of the hadronic mass and charged lepton energy, 
summarised in Table~\ref{tab:4}. 
Owing to the large boost of $B$ hadrons in $Z^0 \to b \bar b$ events, the 
acceptance of the analyses can be extended down to the start of the lepton energy 
spectrum, making their theoretical interpretation more direct. The results correspond 
to $B_d^0$ and $B_u^-$ mesons decays only.

\begin{table}
\caption{\sl Preliminary {\sc Delphi} results for the three leptonic and hadronic 
moments.}
\begin{center}
\begin{tabular}{|l|c c c c|}
\hline 
Moment & Result & (stat) & (syst) & \\
\hline \hline
$M_1(E_{\ell})$  & (1.383 & $\pm$ 0.012 & $\pm$0.009)  & GeV~ \\
$M_2(E_{\ell})$ & (0.192 & $\pm$ 0.005 & $\pm$0.008)  & GeV$^2$\\
$M_3(E_{\ell})$ &(-0.029 & $\pm$ 0.005 & $\pm$0.006)  & GeV$^3$\\ \hline
$M_1(M_X)$         & (0.534 & $\pm$ 0.041 & $\pm$ 0.074) & GeV$^2$ \\
$M_2(M_X)$        & (1.226 & $\pm$ 0.158 & $\pm$ 0.152) & GeV$^4$ \\
$M_3(M_X)$        & (2.970 & $\pm$ 0.673 & $\pm$ 0.478) & GeV$^6$ \\ 
\hline
\end{tabular}
\end{center}
\label{tab:4}
\end{table}

We perform a $\chi^2$ fit to these six moments, using the two theoretical frameworks 
discussed above. In the fit we also impose some additional constraints derived from 
independent determinations. 
 
In the kinetic mass scheme, we fit the full set of parameters: $m_b(1~{\rm{GeV}})$, 
$m_c(1~{\rm{GeV}})$, $\mu_{\pi}^2$,   $\rho_D^3$ and
$\rho_{LS}^3$. 
We impose  $\mu_G^2$=(0.35$\pm$0.05)~GeV$^2$
\cite{Uraltsev:2001ih} and 
$\rho_{LS}^3$=(-0.15$\pm$0.15)~GeV$^3$.
Two mass constraints have also been applied:
$m_b(1~{\rm{GeV}})$=(4.57$\pm$0.10)~GeV~\cite{mb}, 
and, to be  conservative,  
$m_c(1~{\rm{GeV}})$=(1.05$\pm$0.30)~GeV.
The most stringent is that on $m_b(1~{\rm{GeV}})$. It must be noted that this 
constraint is largely equivalent to that derived from the first moment of the photon 
energy spectrum in $b \to s \gamma$ in other studies~\cite{Cronin-Hennessy:2001fk}. 
 Results are obtained for $\alpha_s(m_b) = 
0.22 \pm 0.01$ and are shown in Table \ref{tab:5}.
In order to study the effect of the bounds on $m_{b,c}$ introduced, 
the fit has been repeated 
unconstrained. Results are consistent, although the accuracy on the masses 
degrades. In particular we find $m_b(1~{\rm{GeV}})$=(4.61$\pm$0.15)~GeV. 
It is interesting to observe that the mass 
constraints applied are of the scale of the fit 
sensitivity. Also, the central values of the heavy quark 
masses are in good agreement with independent determinations~\cite{mb,melyel}.

In the alternative approach based on pole masses, 
the fit extracts $\bar{\Lambda}$, $\lambda_1$, $\lambda_2$, $\rho_1$
and $\rho_2$. We fix ${\cal{T}}_i = 0$ and impose two constraints from 
$M_{B^*}-M_B$ and $M_{D^*}-M_D$
which effectively reduce by two the number of free parameters. The
results are given in Table~\ref{tab:6}. 

Projections of the constraints from the six moments in the $m_b$-$\mu_{\pi}^2$ and 
$m_b$-$\rho_D^3$ planes are shown in Fig.~1 and those in the 
$\bar{\Lambda}$-$\lambda_1$ and $\bar{\Lambda}$-$\rho_1$ planes in Fig.~2. 
The  $\chi^2/{\rm{n.d.f.}}$ of the fits is  0.96 and 0.35 in the two 
formulations. Since the contributions proportional to $\rho_{LS}^3$ in 
the moment expressions are numerically suppressed, the fit is only marginally sensitive 
to its size and the result is determined by the constraint applied. By removing this, 
the fit would give $\rho_{LS}^3$=(-1.0$\pm$0.7)~GeV$^3$.

In contrast, the value of the leading $1/m_b^3$ correction
(parameterised by $\rho^3_D$ or $\rho_1$) can 
be determined with satisfactory accuracy and its range agrees with
theoretical expectations~\cite{Uraltsev:2001ih}.

\begin{table}
\caption{\sl Results of fit for the $m_b(\mu)$, $m_c(\mu)$ and 
$\mu_{\pi}^2(\mu)$ formalism.}
\begin{center}
\begin{tabular}{|l|c c c c|}
\hline
Fit & Fit & Fit & Syst. & \\
Parameter & Values & Uncertainty & Uncertainty & \\ \hline
$m_b(1~{\rm{GeV}})$ & 4.59 & $\pm$ 0.08 & $\pm$ 0.01 & GeV~\\
$m_c(1~{\rm{GeV}})$ & 1.13 & $\pm$ 0.13 & $\pm$ 0.03 & GeV~\\
$\mu_{\pi}^2(1~{\rm{GeV}})$ & 0.31 & $\pm$ 0.07 & $\pm$ 0.02 & GeV$^2$\\
$\rho_D^3$    & 0.05 & $\pm$ 0.04 & $\pm$ 0.01 & GeV$^3$ \\ \hline
\end{tabular}
\end{center}
\label{tab:5}
\end{table}

\begin{table}
\caption{\sl Results of fit for the $\bar{\Lambda}$-$\lambda_1$ formalism.}
\begin{center}
\begin{tabular}{|l|c c c c|}
\hline
Fit & Fit & Fit & Syst. & \\
Parameter & Values & Uncertainty & Uncertainty & \\ \hline
$\bar{\Lambda}$ & ~0.40 & $\pm$ 0.10 & $\pm$ 0.02 & GeV~\\
$\lambda_1$   & -0.15 & $\pm$ 0.07 & $\pm$ 0.03 & GeV$^2$\\
$\lambda_2$   & ~0.12 & $\pm$ 0.01 & $\pm$ 0.01 & GeV$^2$\\
$\rho_1$      & -0.01 & $\pm$ 0.03 & $\pm$ 0.03 & GeV$^3$ \\ \hline
$\rho_2$      & ~0.03 & $\pm$ 0.03 & $\pm$ 0.01 & GeV$^3$ \\  \hline
\end{tabular}
\end{center}
\label{tab:6}
\end{table}

Systematic uncertainties due to ranges of residual parameters which have been fixed 
and missing terms in the expansions have been estimated. 
For the running mass formalism we propagate the uncertainty on $\alpha_s$ and 
evaluate the effect of removing the BLM corrections from the lepton moments. 
In this scheme that is a small effect and higher order
perturbative corrections are expected to be under control. 
Dimensional estimates suggest that 
$1/m_b^4$ effects do not exceed the present experimental
resolution.  Other 
systematic uncertainties will be addressed in a dedicated  publication.

For the 
$\bar{\Lambda}$-$\lambda_1$ formalism we take the effect of 
${\cal{T}}_i = (0.0 \pm 0.50)^3$~GeV$^3$, $\alpha_s$ = 0.22$\pm$0.01  
and we also estimate the effect of the missing corrections to third moments as
$M_B^6(0.001\pm0.0005)\,\beta_0\,(\alpha_s/\pi)^2$
and $M_B^6(0.003\pm0.003)\,\bar\Lambda/\bar{M}_B\, \alpha_s/\pi$.

The fit was also repeated using only the first two moments, leaving free  
$m_b(1~{\rm{GeV}})$, $\mu_{\pi}^2(1~{\rm{GeV}})$ and $\bar{\Lambda}$, 
$\lambda_1$, respectively. The other parameters were fixed 
to the central values obtained in the full fit. Results agreed with those from the 
full fit. In particular, the values of $\bar{\Lambda} = 0.42 \pm 0.07$(stat.)~GeV and 
$\lambda_1 =(-0.17 \pm 0.05$(stat.))~GeV$^2$ agree with the recent result reported by 
the {\sc Cleo} Collaboration~\cite{Briere:2002hw}, which uses the first moments of the 
charged lepton energy to obtain $\bar{\Lambda} = 0.39 \pm 0.07$~GeV and 
$\lambda_1 =(-0.25 \pm 0.05$)~GeV$^2$.

There are several facets of these results to be looked at. One interesting 
piece of information comes from the correlation between $m_c$ and 
$m_b$ extracted from the fit. It corresponds 
to $m_c(1~{\rm{GeV}}) = 1.63 \times (m_b(1~{\rm{GeV}})-3.91)$. 
Therefore a competitive value of the charm mass can be extracted from a 
precise determination of $m_b$. Using, for instance, 
$m_b(1~{\rm{GeV}}) = (4.60 \pm 0.05)$~GeV would give 
$m_c(1~{\rm{GeV}}) = (1.13 \pm 0.09)$~GeV. This can be 
compared to the present typical lattice uncertainties which range 
between 50 and 120~MeV~\cite{mc2}. 

In the running mass scheme, the expansions of Eq.(\ref{eq:05}), for the 
$B$ and $D$ mesons are not used in the fit.
It is therefore possible to test {\it a posteriori} the consistency of the meson mass 
expansion by comparing the $\bar{\Lambda}$ values obtained in the two cases.
We find $ \bar\Lambda( B) - \bar\Lambda( D) =-0.086\pm$0.092. This is
also a test of the size of the non-local terms.

\begin{figure}
\begin{center}
\begin{tabular}{c c}
\epsfig{file=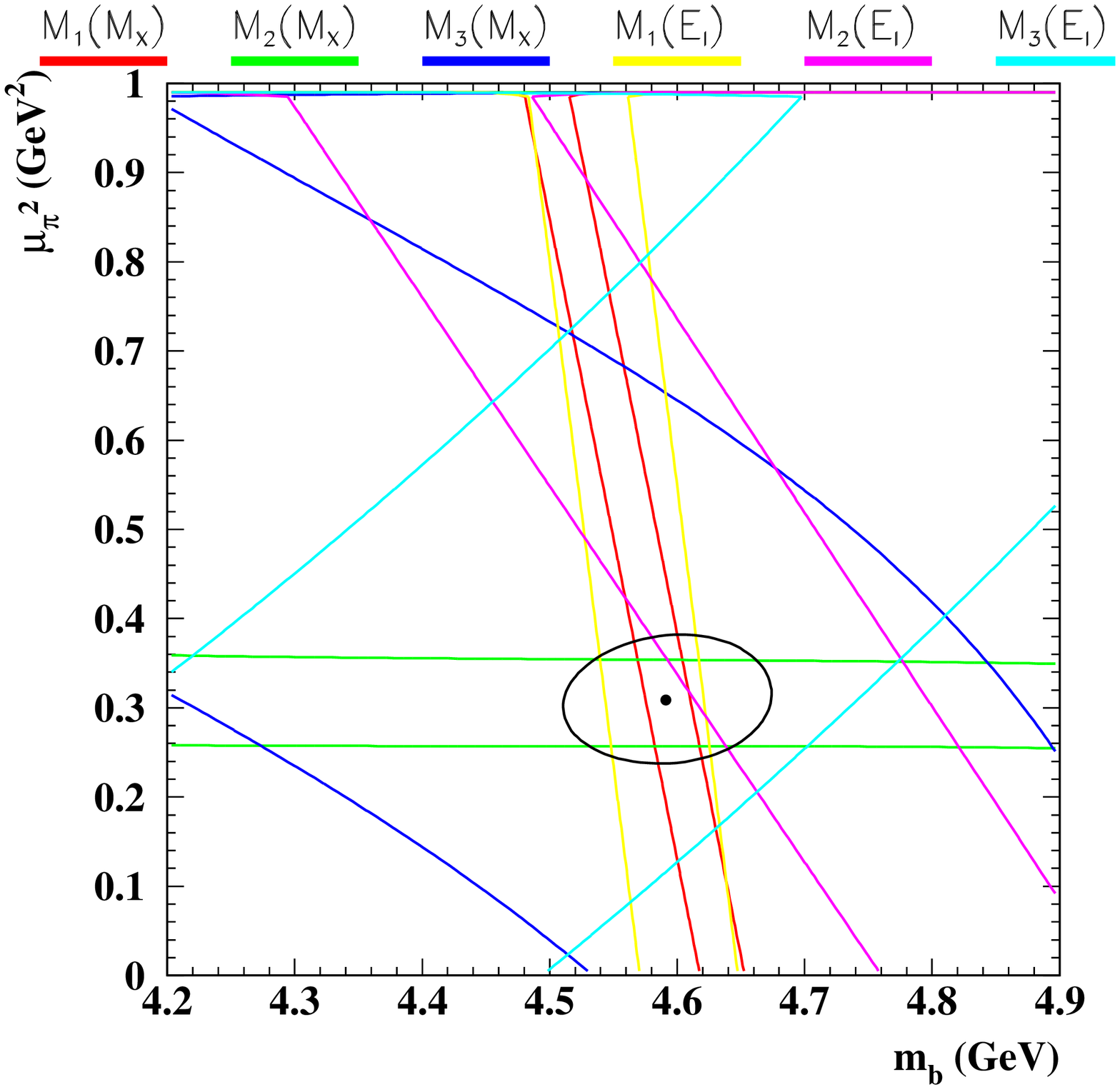,width=8.5cm} &
\epsfig{file=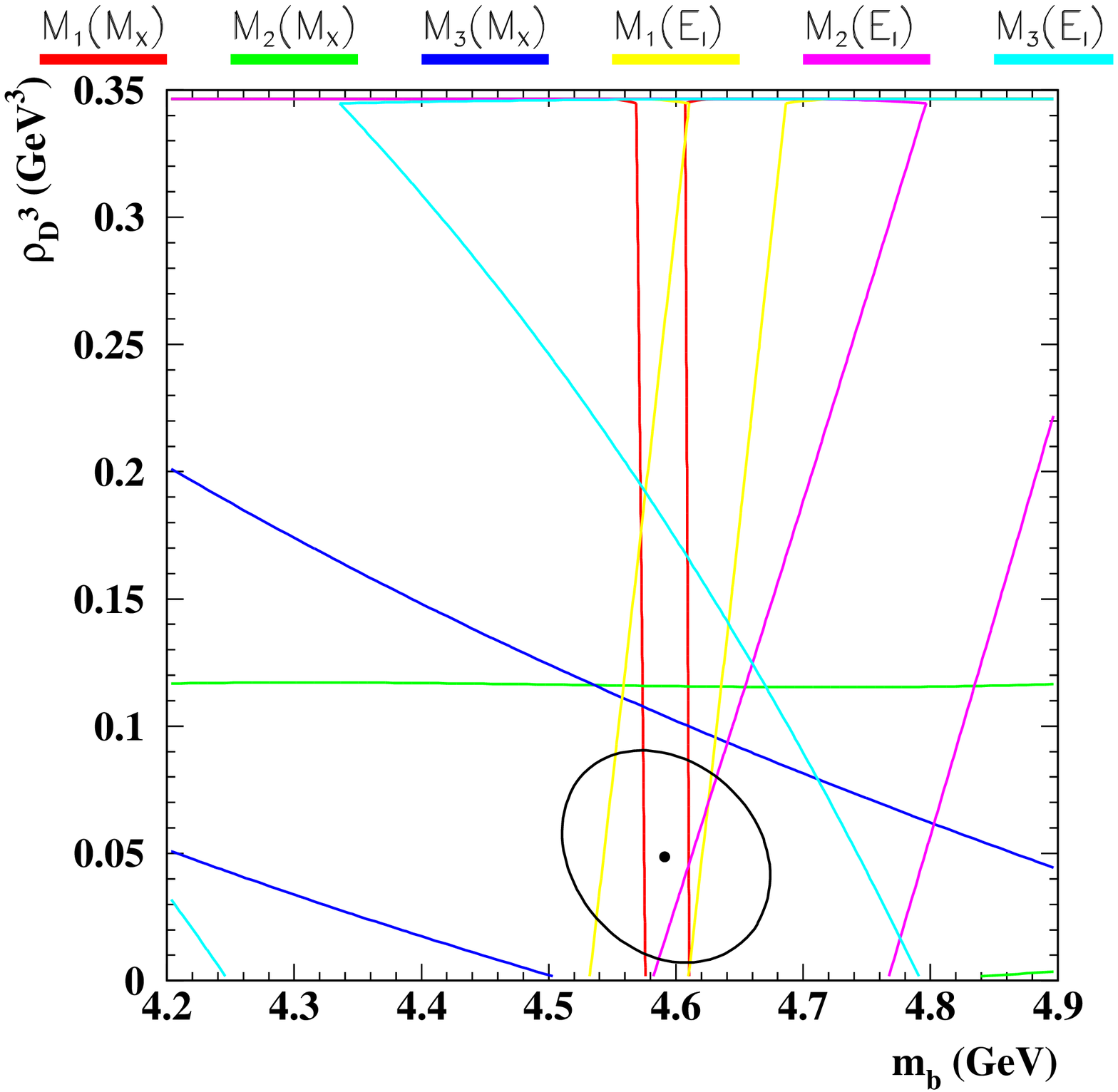,width=8.5cm} \\
\end{tabular}
\caption[]{\sl The projection of the constraints of the six measured moments on the 
$m_b(1~{\rm{GeV}})$-$\mu_{\pi}^2(1~{\rm{GeV}})$ (left) and 
$m_b(1~{\rm{GeV}})$-$\rho_D^3$ (right) planes. The bands correspond to the total 
measurement accuracy and are given by keeping all the other 
parameters at their central values. The ellipses represent the 1~$\sigma$ contours.}
\end{center}
\label{fig:1}
\end{figure}

\begin{figure}
\begin{center}
\begin{tabular}{c c}
\epsfig{file=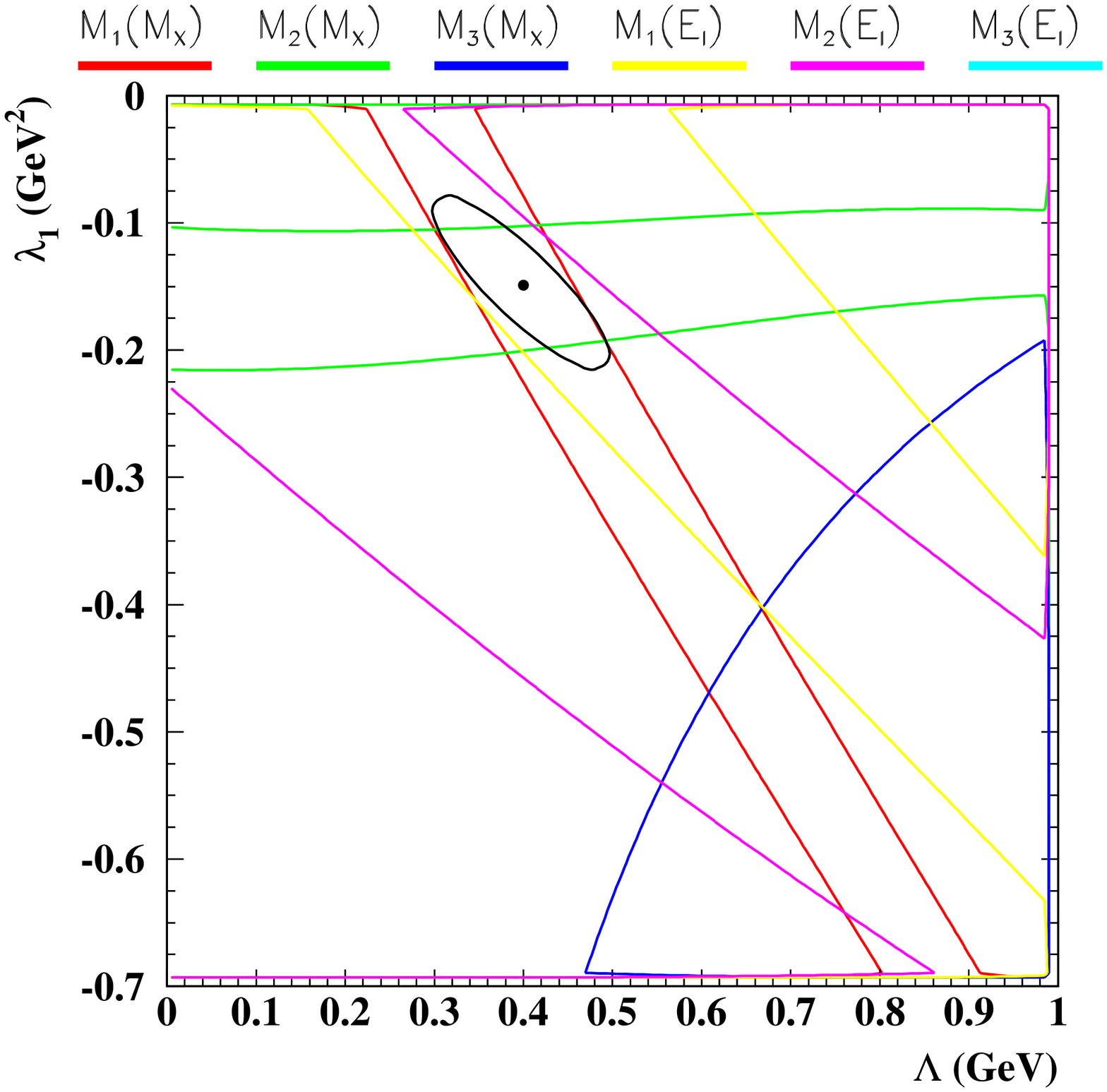,width=8.5cm} &
\epsfig{file=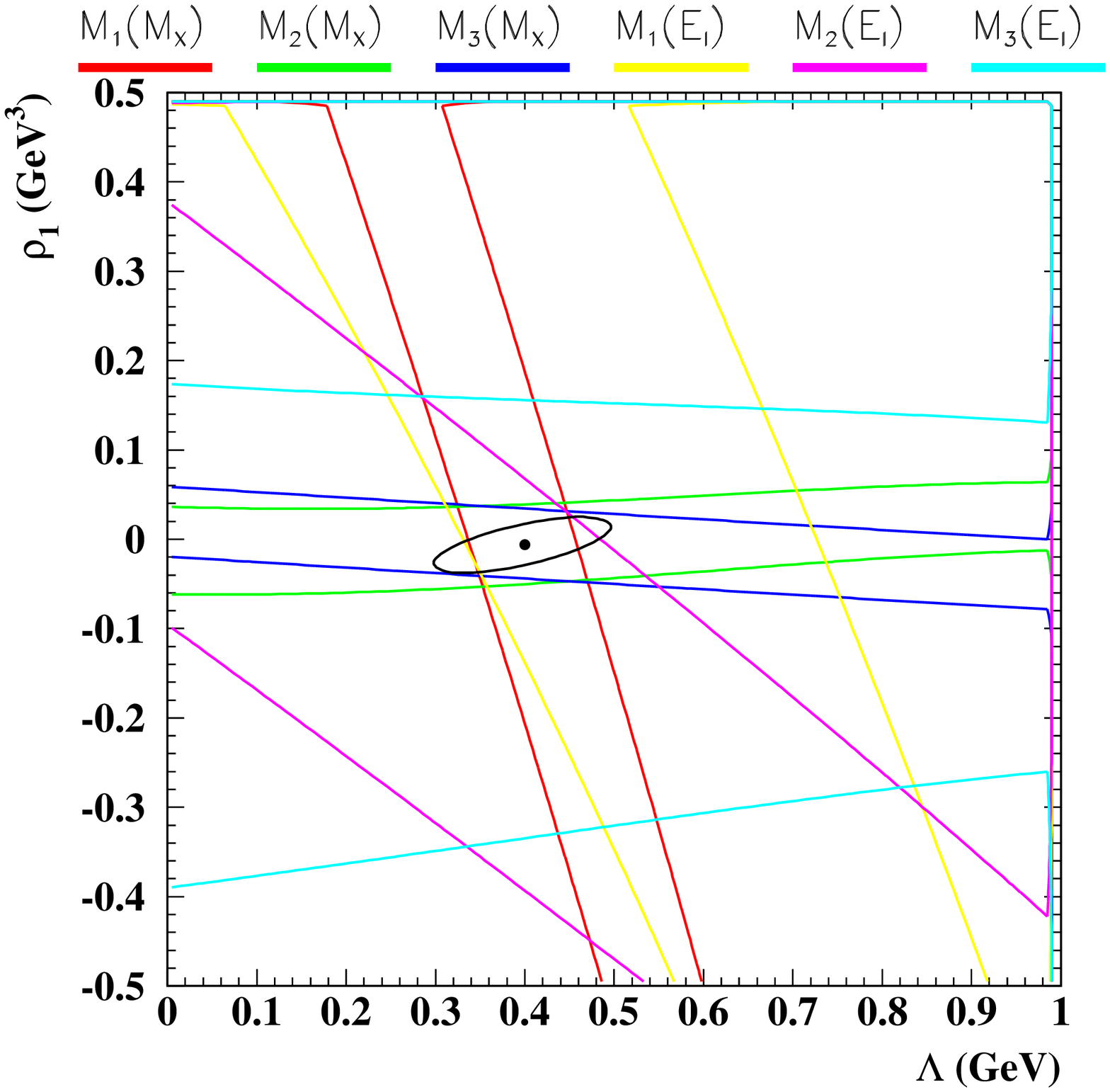,width=8.5cm} \\
\end{tabular}
\caption[]{\sl The projection of the constraints of the six measured moments on the 
$\bar{\Lambda}$-$\lambda_1$ (left) and $\bar{\Lambda}$-$\rho_1$ (right) planes. The 
bands correspond to the total measurement accuracy and are given by keeping all the 
other parameters at their central values. The ellipses represent the 1~$\sigma$ 
contours.}
\end{center}
\label{fig:2}
\end{figure}

In both approaches, the OPE predictions for the six moments, computed
with the  available precision, have a common intersection in the 
multi-parameter space and the quality of the fit is good. 
Within the present experimental accuracy, 
we therefore do not see the need to introduce higher order terms to 
establish agreement with the data. 
In particular, the first leptonic and hadronic moments are highly
correlated and depend on nearly the same combination of heavy quark
masses. Fixing this from $M_1(M_X)$, one finds
$M_1(E_\ell) = 1.377$~GeV which agrees well with the measured value of 
(1.383$\pm$0.015)~GeV. This provides a non-trivial consistency check of the 
OPE. The overall agreement represents both a test of the theory and suggests 
constraints on the size of the $1/m_b^4$ terms and of other missing corrections. 
Similarly, the observed agreement strongly supports the validity of
quark-hadron duality in the $B$ decay shape variables.

At present the achieved experimental resolution matches the available theoretical 
accuracy. With more precise data soon becoming available, it is important to 
improve the latter, particularly for higher hadronic moments. One way to 
improve the convergence of the heavy quark expansion could be to
employ different kinematic variables. We propose to 
consider ${\cal{N}}_{X}^2 = M_X^2 -2 \tilde{\Lambda} E_X$, where $M_X$ and $E_X$ are 
the hadronic mass and energy and $\tilde{\Lambda}$ a fixed mass parameter. Choosing 
$\tilde{\Lambda}$ near $M_B - m_b{\rm{(1~GeV)}} \simeq 0.65$~GeV, suppresses 
terms with $l \ge 1$ in Eq.~(\ref{eq:04}) and results in a better convergence of
higher moments \cite{kolyatalk}.
The use of this variable should be feasible at $B$ factories, where the kinematics 
allows an accurate reconstruction of both the mass and energy of the hadronic 
system in s.l.\ $B$ decays. 

\subsection{Implications for $|V_{cb}|$}
The value of $\Vcb$ obtained from the total s.l.\ decay width depends on the 
OPE parameters extracted above. We discuss now the implications of our results for 
$\Vcb$, using the input parameters given in Table~\ref{tab:meast}. The uncertainties 
on the ${\rm BR}(b \rightarrow X \ell^- \nu)$ have been increased compared to 
ref.~\cite{pdg} for not using the heavy quark forward-backward asymmetries in the 
LEP global electroweak fit and to account for the $\pm15\%$ uncertainty on the 
equality of s.l.\ partial width of $b$ baryons and mesons.

\begin{table}[ht!]
\caption{Input values used for the determination of $|V_{cb}|$.}
\begin{center}
\begin{tabular}{|c|c|}
\hline
Measurement & Value~\cite{pdg} \\
\hline
$b$-hadron lifetime & $(1.564 \pm 0.014)~{\rm ps}$ \\
${\rm BR}(b \rightarrow X \ell^- \nu)$ & $(10.59 \pm 0.31)~\%$ \\
${\rm BR}(b \rightarrow X_u \ell^- \nu)$ & $(0.17 \pm 0.05)~\%$ \\
\hline
\end{tabular}
\label{tab:meast}
\end{center}
\end{table}
The inclusive s.l.\ decay width has been calculated through second order in perturbative 
QCD. Second order BLM corrections were obtained in~\cite{blm}, all-order BLM
terms are available from~\cite{blm2}, whereas second-order non-BLM
corrections have been estimated in~\cite{czarnecki}. Non-perturbative 
corrections start at order ${\cal O}(1/m_b^2)$~\cite{bigi1} and
${\cal O}(1/m_b^3)$ corrections have also been calculated~\cite{mb3}. 
Electroweak corrections have also been taken into 
account~\cite{Sirlin:1981ie}.

The determination of $\Vcb$ and the contributions of the various parameters in 
the kinetic mass scheme is described in~\cite{urabig}. An approximate 
formula which displays the dependence on the different parameters is:
{\small
\begin{eqnarray}
\Vcb = \Vcb_0 \!\!& &\left [ 1 -0.65 \left ( m_b(1) -4.6~\gevcd \right )
+0.40 \left ( m_c(1) -1.15~\gevcd \right ) \right. \nonumber \\
& & +0.01 \left ( \mu_{\pi}^2 -0.4~\gevd \right )
+0.10 \left ( \rho_D^3 -0.12~\gevt \right ) \nonumber \\
& & \left. +0.06 \left ( \mu_G^2-0.35~\gevd \right ) 
-0.01 \left ( \rho_{LS}^3 +0.17~\gevt \right ) \right ].
\label{eq:vcbnum}
\end{eqnarray}
}

A detailed discussion of the theoretical uncertainties on $|V_{cb}|$ goes beyond 
the scope of this paper. Here we focus on the uncertainty arising from 
the heavy quark masses and non-perturbative parameters determined in the fit.
It is evaluated using the full fit error matrix which leads to $\pm1.5\%$. 
There is an additional uncertainty coming from the limited accuracy of the theoretical 
expressions which have been used.
We take the range $m_b/2 < \mu' < m_b$ for the scale $\mu'$ at which 
$\alpha_s$ is evaluated and find a 
$\pm 1\%$ effect\footnote{Incorporating the third-order BLM
correction suppresses this scale dependence. Combining
Refs.~\cite{blm2} and \cite{dipole}, we find the third-order BLM correction
to $\Gamma_{\rm sl}(b\!\to\!c)$ to be $\approx -50(\alpha_s/\pi)^3$ in
this scheme. This increases $|V_{cb}|$ by 1\% for $\mu'\!=\!m_b$, and leaves it 
nearly unchanged, compared to two loops, for $\mu'\!=\!m_b/2$.}.
In summary, we obtain:
\beq
\Vcb = 0.0419 \times \left ( 1 \pm \left. 0.016 \right |_{meas}
\pm \left. 0.015 \right |_{fit} 
\pm \left. 0.010 \right |_{pert}
\right ),
\eeq
where the first uncertainty reflects the accuracy on the s.l.\ width determination. 

The expression for the inclusive $b$ s.l.\ width in the pole mass scheme is known 
to the same order. 
The fit results have been used to obtain:
\beq
\Vcb = 0.0413 \times \left ( 1 \pm \left. 0.016 \right |_{meas}
\pm \left. 0.017 \right |_{fit} 
\pm \left. 0.006 \right |_{nl} 
\pm \left. 0.021 \right |_{pert}
\right).
\eeq
Again, the first two uncertainties correspond to the decay width measurement and to the 
fitted parameters, respectively. The third uncertainty 
refers to the ${\cal T}_{i=1,4}$ parameters which 
have been varied within the range $(0 \pm (0.5)^3)~\gevt$.
The uncertainty from the truncation of the perturbative QCD series is again estimated 
by varying the scale at which $\alpha_s$ is evaluated between $m_b/2$ and $2 m_b$. 
Here the perturbative uncertainty is larger and reflects the slower convergence of the 
perturbative series when the pole mass scheme is employed. 

\section{Conclusions}

The values of the heavy quark masses have been determined, together with the 
leading $1/m_b^2$ and $1/m_b^3$ parameters, from a fit to the first three 
moments of the charged lepton energy and hadronic mass spectra in
s.l.\ decays, measured in a preliminary analysis of
the {\sc Delphi} data. The absence of a charged lepton energy cut in the analysis 
makes the heavy quark expansion more reliable and allows us to include higher 
moments. We have adopted two different formalisms: one based on low-energy 
running quark 
masses, which does not rely on a $1/m_c$ expansion, and the other on pole quark masses. 
The constraints from the six moments agree well and the size of the dominant 
$1/m_b^3$ term has been found to be compatible with theoretical estimates.  The fit 
is largely insensitive to non-local correlators and to the spin-orbital operator.
   
Propagating the ranges of the OPE parameters to the determination of $|V_{cb}|$ 
reduces the theoretical uncertainty due to the $1/m_b^3$ corrections below 2\%.
Furthermore, the use of a fit
changes the nature of these uncertainties and partly removes 
the arbitrariness arising from estimates based on parameters ranges.    

\subsubsection*{Acknowledgements}
We thank D.~Benson, I.~Bigi and Z.~Ligeti for interesting discussions. The work of P.G. 
was supported by a EU Marie Curie Fellowship. The work of N.U. was supported in part by 
the NSF under grant number PHY-0087419.

\subsubsection*{Note}
During the final stage of this work, a new analysis of s.l.\ moments  has 
appeared~\cite{Bauer:2002sh}. There are several differences with our approach, but 
the results are qualitatively consistent with our findings.

\end{document}